\documentclass[epsf,10pt]{article}
\usepackage{amsmath,epsfig}

\begin{document}

%%%%%%%%%%%%%%%%%%%%%%%%%%%%%%% begin of macro %%%%%%%%%%%%%%%%%%%%%%%%%%
\newtheorem{theorem}{Theorem}[section]
\newtheorem{lemma}{Lemma}[section]
\newtheorem{corollary}{Corollary}[section]
\newtheorem{proposition}{Proposition}[section]
\newtheorem{definition}{Definition}[section]
\newcommand{\blackslug}{\penalty 1000\hbox{
    \vrule height 8pt width .4pt\hskip -.4pt
    \vbox{\hrule width 8pt height .4pt\vskip -.4pt
          \vskip 8pt
      \vskip -.4pt\hrule width 8pt height .4pt}
    \hskip -3.9pt
    \vrule height 8pt width .4pt}}
\newcommand{\proofend}{\quad\blackslug}
\newenvironment{proof}{\vspace{1mm} \noindent {\sc Proof.}$\;$\rm}{\qed}
\newcommand{\qed}{\hspace*{\fill}\blackslug}
\def\boxit#1{\vbox{\hrule\hbox{\vrule\kern4pt
  \vbox{\kern1pt#1\kern1pt}
\kern2pt\vrule}\hrule}}
%%%%%%%%%%%%%%%%%%%%%%%%%%%%%%% end of macro %%%%%%%%%%%%%%%%%%%%%%%%%%%%%
% \addtolength{\baselineskip}{+0.6mm}

\title{
An Improved Parameterized Algorithm for the Independent Feedback Vertex Set Problem}
\author{
Yinglei Song\thanks{School of Computer Science and Engineering,
Jiangsu University of Science and Technology,
Zhenjiang, Jiangsu 212003, China.
Email: {\tt yingleisong@gmail.com.}}
}

\date{}
\maketitle

\begin{abstract}
\noindent In this paper, we develop a new parameterized algorithm for the {\sc Independent Feedback Vertex Set} (IFVS) problem. Given a graph $G=(V,E)$, the goal of the problem is to determine whether there exists a vertex subset $F\subseteq V$ such that $V-F$ induces a forest in $G$ and $F$ is an independent set. We show that there exists a parameterized algorithm that can determine whether a graph contains an IFVS of size $k$ or not in time $O(4^kn^{2})$. To our best knowledge, this result improves the known upper bound for this problem, which is $O^{*}(5^{k}n^{O(1)})$.

\end{abstract}


\section{Introduction}

A {\it feedback vertex set} in a graph $G=(V,E)$ is a vertex subset $F \subseteq V$ such that vertices in $V-F$ induce a forest in $G$. Feedback vertex sets have important applications in deadlock recovery in the development of operating systems \cite{silberschatz} and database management systems \cite{molina}. Moreover, $F$ is an {\it independent feedback vertex set} if it induces an independent set in $G$.

In the past two decades, the {\sc Feedback Vertex Set} (FVS) problem has been intensively studied. The goal of the problem is to determine whether a graph contains a FVS of size $k$ or not. In \cite{karp}, it is shown that the problem is NP-complete.
So far, a few exact algorithms have been developed to compute a minimum FVS in a graph. For example, in \cite{razgon}, an algorithm that needs $O(1.9053^{n})$ time was developed to find a minimum FVS in a graph, this algorithm is also the first one that breaks the trivial $O(2^{n}n)$ bound. In \cite{fomin}, an elegant algorithm that can enumerate all induced forests in a graph is used to find the minimum FVS. This algorithm needs $O(1.7548^{n})$ computation time. Recently, the upper bound of this problem is improved to $O(1.7347^n)$ \cite{fomin2}.

Parameterized computation provides another potentially practical solution for problems that are computationally intractable. Specifically, one or a few parameters in some intractable problems can be identified and parameterized computation studies whether efficient algorithms exist for these problems while some or all of the parameters are small. A parameterized problem may contain a few parameters $k_1, k_2, \cdots, k_l$ and the problem is {\it fixed parameter tractable} if it can be solved in time $O(f(k_1, k_2, \cdots, k_l)n^{c})$,where $f$ is a function of $k_1, k_2, \cdots, k_l$, $n$ is the size of the problem and $c$ is a constant independent of all parameters. For example, the {\sc Vertex Cover} problem is to determine whether a graph $G=(V,E)$ contains a vertex cover of size at most $k$ or not. The problem is NP-complete. However, a simple parameterized algorithm can solve the problem in time $O(2^{k}|V|)$ \cite{downey}. In practice, this algorithm can be used to efficiently solve the {\sc Vertex Cover} problem when the parameter $k$ is fixed and small. On the other hand, some problems do not have known efficient parameterized solutions and are therefore parameterized intractable. Similar to the conventional complexity theory, a hierarchy of complexity classes has been constructed to describe the parameterized complexity of these problems \cite{downey}. For example,
the {\sc Independent Set} problem is to decide whether a graph contains an independent set of size $k$ or not and has been shown to be W[1]-complete \cite{downey2}. It cannot be solved with an efficient parameterized algorithm unless all problems in W[1] are fixed parameter tractable. A thorough investigation on these parameterized complexity classes are provided in \cite{downey}.

The parameterized FVS problem uses the size of an FVS as the parameter $k$. In \cite{bodlaender,dom,downey,downey2,dehne,guo,guo2,kanj,raman,raman2}, it is shown that the problem is fixed parameter tractable and a few algorithms that can solve the problem in time $O(2^{O(k)}n^{O(1)})$ are developed. In \cite{chen}, it is shown that the problem can be solved in time $O(5^{k}kn^{2})$. Recently, the upper bound of this problem is further improved to $O(3.83^{k}kn^{2})$ \cite{cao}.

Although the FVS problem has attracted a lot of attention from researchers in computer science,
little effort has been made toward the IFVS problem. Recently,
it is shown that the FVS problem can be reduced to the IFVS problem in
polynomial time while preserving the parameter $k$ \cite{misra}. The reduction is
simply subdividing each graph edge once. The IFVS problem is
thus more general than the FVS problem. In \cite{misra}, it is shown that the
IFVS problem can be solved in time $O^{*}(5^kn^{O(1)})$.

In this paper, we develop a new parameterized algorithm that can determine whether a graph contains an IFVS of size $k$ or not in time $O(4^{k}n^2)$. Our algorithm follows the {\it iterative compression} approach developed in \cite{reed}. This technique has been used to develop parameterized algorithms for many problems \cite{dehne,dom,guo2}. Specifically, given a graph $G$, the process starts with an empty graph and adds vertices one by one to reconstruct $G$. When a vertex is added, edges are also added to join the vertex with its neighbors that have been included in the graph. A series of graph $G_0,G_1,\cdots,G_n$ can thus be constructed during the process, where $G_0$ is an empty graph and $G_n=G$. Starting with $G_{2}$, the approach processes each $G_i$ ($2 \leq i \leq n$) to determine whether it contains an IFVS of size $k$ or not. If the answer is ``yes'', the approach finds such a set $F_i$  in $G_i$ and continues to process $G_{i+1}$, otherwise the approach outputs ``no'' since $G$ does not contain an IFVS of size $k$ if one of its subgraph does not contain such an IFVS. The approach outputs ``yes'' if eventually $G$ is found to contain an IFVS of size $k$. A detailed description of the approach can be found in \cite{reed} and some other related work such as \cite{dehne,dom,guo2}.

It is not difficult to see that, the key part of the above approach is to develop an algorithm that can determine whether $G_{i+1}$ contains an IFVS of size $k$ or not, given an FVS of size $k+1$. The FVS can be obtained by including all vertices in $F_i$ and the additional vertex that is added to $G_i$ to create $G_{i+1}$. The major contribution of this paper is the development of such an algorithm. Other than using a branching approach, we develop a dynamic programming based algorithm that can accomplish the task in time $O(4^{k}n)$. Since the graph contains $n$ vertices in total, the total amount of computation time needed by the approach is thus at most $O(4^{k}n^2)$. Based on the simple parameter preserving polynomial time reduction that reduces the FVS problem to IFVS problem, we also immediately obtain a parameterized algorithm that solves the FVS problem in time $O(4^{k}n^2)$.

\section{Notations and Preliminaries}

The graphs in this paper are undirected graphs without loops.
For a given graph $G=(V,E)$ and a vertex subset $V' \subseteq V$, $G[V']$ is the subgraph induced on the vertices in $V'$. A vertex $u$ is a {\it neighbor} of $V'$ if it is joined to one of the vertices in $V'$ by an edge in $G'$. We use $N_{G}[V']$ to denote the set of neighbors of $V'$ in $G$. We may omit the subscript $G$ when the underlying graph is clear in the context.
Given two graphs $G_1=(V_1, E_1)$ and $G_2=(V_2, E_2)$, we use
$G_1 \cup G_2$ to denote the graph $(V_1 \cup V_2, E_1 \cup E_2)$. Furthermore, to simplify the notation, we use $G-V'$ to denote subgraph $G[V-V']$. A vertex subset $V'$ is a {\it feedback vertex cover} in $G$ if $G-V'$ is a forest. The objective of our algorithm is to find a feedback vertex cover that contains minimum number of vertices.

\section{The Algorithm}

Our algorithm is based on the following Lemma.
\begin{lemma}
\label{lm1}
\rm
Given a graph $G=(V,E)$ and an FVS $F$ in $G$, a minimum
IFVS in $G$ can be computed in time $O(4^{|F|}|V|)$.

\begin{proof}
Since $F$ is an FVS, $G[V-F]$ is a forest. As the first step of the algorithm, we arbitrarily choose a vertex in each tree in $G[V-F]$ as the root of the tree. We color each vertex in $V-F$ with black. We then check the number of children of each vertex in the tree and find those that have more than two children. For each such vertex $u$, we remove the edges that join it with its children $c_1,c_2,\cdots, c_d$, where
$d$ is the number of children in the tree. We then create a path of white vertices $w_1,w_2,\cdots,w_{d-1}$. $v_{d-1}$ is joined to $u$ with an edge. To construct a binary tree, $w_1$ is joined to $c_1$ and $c_2$ with two edges and each white vertex $w_t$  ($2 \leq t \leq d-1$) in the path is joined to $c_{t+1}$ with an edge. Each white vertex $w_t$ is said to be {\it equal} to $u$. We use $H$ to denote the resulting forest. The number of white vertices in $H$ is at most
\begin{equation}
\sum_{i=1}^{|V|-|F|}{d_i}\leq 2(|V|-|F|)
\end{equation}
where $d_i$'s are the degrees of all vertices in $G[V-F]$. The number of vertices in $H$ is thus at most $3(|V|-|F|)$. Figure \ref{fig1} illustrates the approach the algorithm uses to convert $T_i$ into a binary tree.

\begin{figure}[!tcp]
\begin{center}
\includegraphics[width=10.0cm, height=5.8cm]{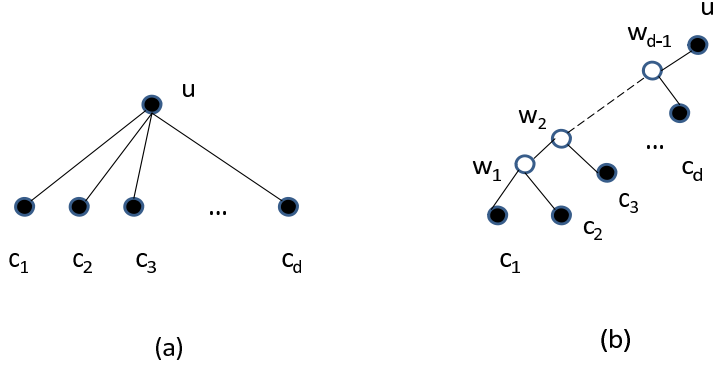}
\caption{The conversion the algorithm uses to convert a tree in $G[V-F]$ into a binary tree.
(a)An internal node $u$ that has $d$ children $c_1,c_2,\cdots,c_d$; (b) A path of
$d-1$ white vertices $w_1,w_2,\cdots,w_{d-1}$ are added to convert the tree into a binary tree.}
\label{fig1}
\end{center}
\end{figure}

In order to find a minimum IFVS $F'$, we consider the vertices in $F' \cap F$ first. We enumerate all subsets in $F$.
For each such subset $S$, we check whether $S$ is an independent set and $G[F-S]$ is acyclic, if it is not the case, $S$ cannot be $F' \cap F$ and we can continue to the next subset. Otherwise, we need to find a minimum subset $S' \subseteq G-F$ such that $S' \cup S$ form an IFVS in $G$.

Without loss of generality, we assume $G[F-S]$ contains $l$ connected components $C_1, C_2, \cdots, C_l$ and let
$C=\{C_1,C_2,\cdots,C_l\}$. Since $H$ is a forest, for each tree $T_i \in H$, we use a dynamic programming algorithm to find a minimum subset $S'_{i}$ in $T_i$ such that $(T_i -S'_{i}) \cup (F-F')$ is acyclic.
We then obtain $S'$ by taking the union of $S'_i$'s for all
$T_i$'s in $G-F$.

We use $r_i$ to denote the root of $T_i$. The algorithm starts with the leaves of the $T_i$ and follows a bottom-up order to scan through all vertices in $T_i$. For a tree node $u$ in $T_i$ and a set $S_u \subseteq C$, we consider the following two cases.
\begin{enumerate}
\item{$u$ is not in the IFVS, there exists a path that connects $u$ to each component in $S_u$ in the subtree rooted at $u$ in $T_i$. Moreover, none of the vertices in the path are in the IFVS,}
\item{$u$ is in the IFVS.}
\end{enumerate}
The algorithm maintains dynamic programming table $M[u, S_u]$ for case 1 and table $P[u]$ for case 2. The two tables store the minimum number of vertices that can be included in the IFVS in the subtree rooted at $T_u$.

For a leaf node $v$ of $T_i$ and a subset $S_v$ of connected components in $G[V-F]$, the values of $M[v, S_v]$ are
initialized to be $0$ if $S_v$ is the set of components that are directly connected to $v$ with an edge. Otherwise, $M[v,S_v]$ is set to be $|V|+1$. If $v \in N_{G}(S)$, $v$ is not in the IFVS, we thus set $P[v]$ to be $|V|+1$, otherwise $P[v]$ is set to be $1$.

For an internal node $u$ in $T_i$, since $T_i$ is a binary tree,
we use $c_1,c_2$ to denote its children in $T_i$. In the case where $u \in N_{G}(S)$, $u$ cannot be in the IFVS and we thus set $P[u]$ to be $|V|+1$. Otherwise, both $M[u, P[u]]$ and $P[u]$ need to be determined with the recursion relations of the dynamic programming.

Next, we determine the recursion relations of the dynamic programming. For a black vertex $u$, we use $W_u$ to denote the set of connected components in $G[F-S]$ that are directly connected to $u$ with an edge. If $W_u$ contains a component that is not in $S_u$, $M[u, S_u]$ is set to be $|V|+1$, since $u$ is connected to any component in $W_u$ regardless of its child nodes in $T_i$. For a white vertex $u$, $W_u=W_b$, where $b$ is the black vertex that is equal to $u$ in $T_i$. The algorithm thus only needs to compute the values of $M[u, S_u]$ for these $S_u$'s that are supersets of $W_u$.

We now assume $W_u \subset S_u$ and $u$ has two children. Based on the color of $u$, we use the following recursion relations to compute $M[u, W_u]$ and $M[u, S_u]$, if $u$ is a black vertex, the recursion relation to compute $M[u,S_u]$ are
\begin{equation}
M[u, W_u] =\min{\{M[c_1,\epsilon]+M[c_2,\epsilon],M[c_1,\epsilon]+P[c_2], P[c_1]+M[c_2,\epsilon],P[c_1]+P[c_2]\}}
\end{equation}
\begin{equation}
M[u, S_u] = \min{\{\min_{S_{c_1},S_{c_2}}{\{M[c_1, S_{c_1}]+M[c_2,S_{c_2}]\}},
                M[c_1,S_{u}-W_{u}]+P[c_2], P[c_1]+M[c_2,S_{u}-W_{u}]\}}
\end{equation}
where $S_{c_1}$ and $S_{c_2}$ are disjoint subsets of $S_u$ and
$S_u-W_u=S_{c_1}\cup S_{c_2}$. The algorithm exhaustively enumerates all such subsets and computes the minimum value of $M[c_1,S_{c_1}]+M[c_2,S_{c_2}]$ over all possible subsets $S_{c_1}$ and $S_{c_2}$.

If $u$ is a white vertex and both $c_1$ and $c_2$ are black vertices, the recursion relation to compute $M[u,S_u]$ is the same as above. If $u$ is white and one of $c_1,c_2$ is white, without loss of generality, we assume $c_1$ is white. Since $u$ and $c_1$ represents the same vertex in $G$, if $u$ is not in the IFVS, $c_1$ cannot be in the IFVS. The recursion relation is thus
\begin{eqnarray}
M[u, W_u] &=& \min{\{M[c_1,W_u]+M[c_2,\epsilon],M[c_1,W_u]+P[c_2]\}} \\
M[u, S_u]&=& \min{\{\min_{S_{c_1},S_{c_2}}{\{M[c_1, S_{c_1}\cup W_u]+M[c_2,S_{c_2}]\}},M[c_1,S_{u}]+P[c_2]\}}
\end{eqnarray}
where $S_{c_1}$ and $S_{c_2}$ are disjoint subsets of $S_u$ and $S_{c_1} \cup S_{c_2}=S_u-W_u$.

If $u$ is black and $u\not\in N_G(S)$, since none of $c_1$ and $c_2$ are included in the IFVS, the recursion relation to compute $P[u]$ is as follows.
\begin{equation}
P[u]=\min_{S_u}{\{M[c_1, S_u]\}}+\min_{S_u}{\{M[c_2,S_u]\}}+1
\end{equation}
where the minimum is taken over all possible subsets $S_u$ of $C$.

If $u$ is white, $u\not\in N_G(S)$, and both $c_1$ and $c_2$ are black, the recursion for computing $P[u]$ is the same as above. If $u$ is white and one of $c_1,c_2$ is white, without loss of generality, we assume $c_1$ is white.
Since $u$ and $c_1$ represents the same vertex and $c_2$ is not included in the IFVS,
the recursion relation is
\begin{equation}
P[u]=P[c_1]+\min_{S_u}{\{M[c_2,S_u]\}}
\end{equation}
where the minimum is taken over all possible subsets $S_u$ of $C$.

If $u$ has only one child, $u$ must be black and we use $c$ to denote its child. If $c$ is black, we use the following recursion relations to compute $M[u, W_u]$, $M[u,S_u]$, and $P[u]$.
\begin{eqnarray}
M[u,W_u] &=& \min{\{M[c,\epsilon],P[c]\}} \\
M[u,S_u] &=& M[c, S_u-W_u] \\
P[u] &=& \min_{S_u}{\{M[c,S_u]\}}+1
\end{eqnarray}
where the minimum is taken over all possible subsets $S_u$ of $C$.

If $c$ is white, $c$ is equal to $u$, we use the following recursion relations to compute $M[u, S_u]$ and $P[u]$.
\begin{eqnarray}
M[u,S_u] &=& M[c,S_u] \\
P[u] & = & P[c]
\end{eqnarray}

After the dynamic programming tables are filled by the algorithm, for each $T_i$($1 \leq k \leq l$), we can find the minimum of $P[r_i]$,  $M[r_i, S_{r_i}]$ for all subsets $S_{r_i}$'s of $C$.We then follow a top-down trace back procedure to find $S'_i$. Note that the above recursions split $S_u$ into two disjoint subsets $S_{c_1}$ and $S_{c_2}$, which can guarantee that no cycles are formed if $u$ is not included in the IFVS. The correctness of this algorithm can thus be easily proved with the principle of induction and its recursion relations.

Before the dynamic programming starts, the algorithm needs to construct $H$ from the trees in $G[V-F]$, it needs at most $O(|V|)$ time. In addition, the algorithm also needs to compute
$W_u$ for each internal node in the trees in $H$ and $N_G(S)$. However, since $S$ contains at most $|F|$ vertices, this needs at most $O(|F||V|)$ time.

During the dynamic programming process, for a given node $u$ and
$S_u$, the algorithm needs to consider all possible disjoint pairs of subsets $S_{c_1}$ and $S_{c_2}$, such that $S_{c_1}\cup S_{c_2} =S_u-W_u$. We denote $t=|S_u|$, the number of such pairs is at most $2^{t}$. The total amount of computation time needed by the dynamic programming on one single vertex in $T_i$ is thus at most
\begin{equation}
O(\sum_{t=0}^{l}{2^{t}\binom{l}{t}})=O(3^{l})
\end{equation}
where the right hand side of the equation can be directly obtained with the binomial theorem.

Since $H$ contains at most $3(|V|-|F|)$ vertices, the dynamic programming algorithm can find $S'$ in time $O(3^{l}|V|)$.For each $S \subseteq F $ such that
$G[F-S]$ is acyclic,  we can find a feedback vertex cover $S \cup S'$. The one that contains the minimum
number of vertices is a minimum IFVS. The total amount of the computation time needed by the algorithm is thus at most
\begin{equation}
O(\sum_{l=1}^{|F|}{3^{l}\binom{|F|}{l}|V|})=O(4^{|F|}|V|)
\end{equation}
The right hand side of the equation again is the direct result of applying the binomial theorem to the left hand side. The theorem has been proved.
\end{proof}
\end{lemma}

Based on the algorithm we have developed in Lemma \ref{lm1}, we can use the iterative compression technique developed in \cite{reed} to develop a parameterized algorithm that can determine whether a graph contains an IFVS of size $k$ in time $O(4^kn^2)$. We thus have the following theorem.

\begin{theorem}
\label{the1}
\rm
Given a graph $G=(V,E)$ and a positive integer parameter $k$, there exists a parameterized algorithm that can determine whether $G$ contains an IFVS of size $k$ or not in time $O(4^k|V|^2)$.

\begin{proof}
We use the iterative compression technique to develop the algorithm. Let $n=|V|$, the algorithm can be sketched as follows.
\begin{enumerate}
\item{Starting with an empty graph $G_0$, add vertices one by one to reconstruct graph $G$, when a vertex $u_i$ is added to $G_i$ ($0 \leq i <n $), edges are added to $G_i$ to join $u_i$ with its neighbors in $G$ that have been included in $G_i$ to construct $G_{i+1}$. We thus can obtain a series of graphs $G_0,G_1,\cdots, G_{n}$, where $G_0$ is an empty graph and $G_{n}=G$;}
\item{Arbitrarily choose one vertex from $G_{2}$ and include it
in set $F_2$;}
\item{For each $i=2$ to $n-1$, given $F_i \cup \{u_i\}$, use the algorithm presented in the proof of Lemma \ref{lm1} to compute a minimum
IFVS $F_{i+1}$ in $G_{i+1}$, if $F_{i+1}$ contains more than $k$
vertices, return ``no'';}
\item{return ``yes''.}
\end{enumerate}

For computation time, step 1 can be accomplished in time $O(n^2)$. From Lemma \ref{lm1}, the computation time needed by step 3 is at most
\begin{equation}
\sum_{i=2}^{n}{O(4^{k}i)}=O(4^{k}n^2)
\end{equation}
The theorem has been proved.
\end{proof}
\end{theorem}

In \cite{misra}, it is shown that the FVS problem can be reduced to the IFVS problem through a simple polynomial time reduction that also preserves the parameter $k$. Theorem \ref{the1} thus implies there exists a dynamic programming based algorithm that solves the FVS problem in time $O(4^kn^2)$. In fact, it is not difficult to see that the algorithm presented in the proof of Lemma \ref{lm1} can be
slightly changed to solve the FVS problem.
\section{Conclusions}

In this paper, we develop a new parameterized algorithm that can determine whether
a graph contains an IFVS of size $k$ or not in time $O(4^{k}n^2)$. To our best knowledge, this result improves the previously known upper bound for this problem, which is $O^*(5^kn^{O(1)})$.Our technique combines a dynamic programming approach with the recently developed iterative compression technique, which has been proved to be useful for developing parameterized algorithms for many NP-complete problems.

Our algorithm does not employ the method of branching, which is an often used approach for developing exact and parameterized algorithms. Indeed, the IFVS is a subset with certain global structural properties. In general,
a branching algorithm employs the local structure of a graph and can hardly be used to further improve
the upper bound of the problem. However, we are still looking for the possibility to combine our method and the
branching approach to further improve
the upper bound of this problem. In fact, recent work \cite{fomin2} has shown that this combination can lead to improved
upper bound results for many NP-hard graph optimization problems.

\end{document}